\documentclass[a4paper,aps,pre,twocolumn,floats,superscriptaddress,showpacs]{revtex4}

\usepackage{epsfig}
\usepackage{bm}
\usepackage{latexsym}
\usepackage{amsmath}

\begin{document}

\newcommand{\avk}{\langle k \rangle}
\newcommand{\fluck}{\langle k^2 \rangle}
\newcommand{\bark}{\bar{k}_{nn}}
\newcommand{\barc}{\bar{c}}

\title{Rate equation approach for correlations in growing network
  models}

\author{Alain Barrat}

\affiliation{Laboratoire de Physique Th\'eorique (UMR du CNRS 8627),
  B\^atiment 210, Universit\'e de Paris-Sud 91405 Orsay, France}

\author{Romualdo Pastor-Satorras}
 
\affiliation{Departament de F\'\i sica i Enginyeria Nuclear, Universitat
  Polit\`ecnica de Catalunya, Campus Nord B4, 08034 Barcelona, Spain}

\date{\today}

\begin{abstract}
  We propose a rate equation approach to compute two vertex
  correlations in scale-free growing network models based in the
  preferential attachment mechanism. The formalism, based on previous
  work of Szab\'o \textit{et al.} [Phys. Rev. E \textbf{67} 056102
  (2002)] for the clustering spectrum, measuring three vertex
  correlations, is based on a rate equation in the continuous degree
  and time approximation for the average degree of the nearest
  neighbors of vertices of degree $k$, with an appropriate
  boundary condition. We study the properties of both two and three
  vertex correlations for linear preferential attachment models, and
  also for a model yielding a large clustering coefficient.  The
  expressions obtained are checked by means of extensive numerical
  simulations.  The rate equation proposed can be generalized to more
  sophisticated growing network models, and also extended to deal with
  related correlation measures.  As an example, we consider the case
  of a recently proposed model of weighted networks, for which we are
  able to compute a weighted two vertex correlation function, taking
  into account the strength of the interactions between connected
  vertices.
\end{abstract}


\maketitle

\section{Introduction}

Many natural and man-made complex systems can be fruitfully
represented and studied in terms of networks or graphs
\cite{bollobas98}, in which the vertices stand for the elementary
units that compose the system, while the edges picture the
interactions or relations between pairs of units.  This topological
representation has found many applications in fields as diverse as the
Internet \cite{romuvespibook}, the World-Wide Web \cite{www99},
biological interacting networks \cite{wagner01,jeong01,maslov02} or
social systems \cite{wass94}, leading to the development of a new
branch of statistical mechanics, the modern theory of complex networks
\cite{barabasi02,mendesbook}.

The empirical study of real complex networks, promoted by the recent
accessibility to computers powerful enough to deal with very large
databases, has uncovered the presence of some typical characteristics.
The three most relevant of these are: (i) The small-world property
\cite{watts98}, defined by an average shortest path length---average
distance between any pair of vertices---increasing very slowly
(logarithmically or slower \cite{havlin03}) with the network size $N$.
(ii) The presence of a large transitivity \cite{wass94}, which implies
that two neighbors of a given vertex are also connected to each other
with large probability.  Transitivity can be quantitatively measured
by means of the clustering coefficient $c_i$ of vertex $i$
\cite{watts98}, defined as the ratio between the number of edges $m_i$
existing betwen the $k_i$ neighbors of $i$, and its maximum possible
value, i.e.  $c_i = 2 m_i / (k_i(k_i-1))$. The average clustering
coefficient, defined as $C = \sum_i c_i /N$, usually takes quite large
values in real complex networks.  (iii) A scale-free behaviour for the
degree distribution $P(k)$ \cite{barabasi02,mendesbook}, defined as
the probability that a vertex is connected to $k$ other vertices (has
degree $k$), that shows a power-law behavior
\begin{equation}
  P(k) \sim k^{-\gamma},
\end{equation}
where $\gamma$ is a characteristic degree exponent, usually in the range
$2< \gamma < 3$.  A major role is especially played by the scale-free
nature of many real complex networks, which implies a large
connectivity heterogeneity, at the basis of the peculiar behavior shown
by dynamical processes taking place on top of these networks, such as
the resilience to random damage \cite{barabasi00,newman00,havlin00},
the spreading of infectious agents
\cite{pv01a,pv01b,lloydsir,sievolution}, or diffusion-annihilation
processes \cite{originalA+A,michelediffusion}.

>From a theoretical point of view, the empirical research has inspired
the proposal of new network models, aimed at reproducing and
explaining the properties exhibited by complex networks. In this
respect, many efforts have been devoted to develop models capable to
account for a scale-free degree distribution. Classical network
modeling was previously based on the Erd\"os-Renyi random graph model
\cite{erdos59,bollobas}, which is a static model (i.e. defined for a
fixed number of vertices $N$) yielding small-world networks with a
Poisson degree distribution. A change of perspective in nework
modeling took place after the introduction of the preferential
attachment paradigm first proposed by Barab\'asi and Albert (BA)
\cite{barab99}. The insight behind this concept is the realization of
two facts: Firstly, most complex networks are the result of a growth
process, in which new vertices are added in time to the system.
Secondly, new edges are not placed at random, but tend to connect to
vertices which already have a large degree. It turns out that these
two ingredients are able to reproduce scale-free degree distributions
with a tunable degree exponent~\cite{barab99,mendes99}.  Moreover, it
has been shown that not all sorts of preferential attachment are able
to generate a power law degree distribution, but only those in which
new edges attach to vertices with a probability strictly linear in
their degree \cite{krap00}, and that some alternative mechanisms, such
as the copying model \cite{kumar00} implicitly define a linear
preferential attachment dynamics.

While a proper characterization and understanding of the origin of the
scale-free degree distribution displayed by most real complex networks
is a fundamental task, it has been recently realized that this
property does not provide a suffient characterization of natural
networks. In fact, these systems seem to exhibit also ubiquitous
degree correlations, which translate in the fact that the degrees of
the vertices at the end points of any given edge are not independent
\cite{alexei,alexei02,assortative,newmanmixing}.  Two vertex degree
correlation can be conveniently measured by means of the conditional
probability $P(k'|k)$ that a vertex of degree $k$ is connected to a
vertex of degree $k'$. For uncorrelated networks, in which this
conditional probability is independent of $k$, it can be estimated as
the probability that any edge end points to a vertex of degree $k'$,
which is simply given by $P_\mathrm{n.c.}(k'|k) = k' P(k') / \avk$
\cite{hiddenvars}.  The empirical evaluation of $P(k'|k)$ in real
networks is usually a cumbersome task, restricted by finite size data
yielding noisy results. For this reason, it is more practical to
analyze instead the average degree of the nearest neighbors of the
vertices of degree $k$, which is formally defined as~\cite{alexei}
\begin{equation}
  \bark(k) = \sum_{k'} k' P(k'|k).
  \label{eq:8}
\end{equation}
For uncorrelated networks, in which $P(k'|k)$ does not depend on $k$,
we have
\begin{equation}
  \bark^\mathrm{n.c.}(k) =    \frac{\fluck}{\avk},
  \label{eq:13}
\end{equation}
independent of $k$.  Thus, a $\bark(k)$ function with an explicit
dependence on the degree signals the presence of two vertex degree
correlations in the network. When $\bark(k)$ is an increasing function
of $k$, the network shows \textit{assortative mixing}
\cite{assortative} (vertices of large degree connected more preferably
with vertices of large degree, and vice-versa). Negative correlations
(low degree vertices connected preferably with large degree vertices),
on the other hand, give rise to \textit{dissasortative mixing},
detected by a decreasing $\bark(k)$ function.
 
Analogously to two vertex correlations, correlations implying three
vertices can be mesured by means of the probability $P(k', k''|k)$
that a vertex of degree $k$ is simultaneously connected to vertices of
degree $k'$ and $k''$. Again, the difficulties in directly estimating
this conditional probability can be overcome by analyzing the
clustering coefficient. The average clustering coefficient of the
vertices of degree $k$ (the clustering spectrum), $\barc(k)$
\cite{alexei02,ravasz02}, can be formally computed as the probability
that a vertex of degree $k$ is connected to vertices of degree $k'$
and $k''$, and that those two vertices are at the same time joined by
an edge, averaged over all the possible values of $k'$ and $k''$
\cite{hiddenvars}. Thus, we can write $\barc(k)$ as a function of the
three vertex correlations as
\begin{equation}
  \barc(k) = \sum_{k', k''} P(k', k''|k) p^k_{k', k''},
\end{equation}
where $p^k_{k', k''}$ is the probability that vertices $k'$ and $k''$
are connected, provided that they have a common neighbor $k$
\footnote{Note that the probability $p^k_{k', k''}$ can depend on the
  degree $k$ of the common vertex}. From this expression, the average
clustering coefficient can be computed as
\begin{equation}
  C = \sum_k P(k)  \barc(k).
\end{equation}
For uncorrelated networks, we have that $P_\mathrm{n.c.}(k',
k''|k)=P_\mathrm{n.c.}(k'|k) P_\mathrm{n.c.}(k''|k)$
\cite{hiddenvars}, and $p^k_{k', k''} = (k'-1)(k''-1)/ \avk N$
\cite{newmanrev}. Therefore we obtain
\begin{equation}
  \barc_\mathrm{n.c.}(k) = \frac{(\fluck - \avk)^2}{ \avk^3 N}.
\end{equation}
That is, $\barc_\mathrm{n.c.}(k)$ is independent of $k$ and equal to
the average clustering coefficient $C$ \cite{newmanrev}. A functional
dependence of $\barc(k)$ on the degree can thus be attributed to the
presence of a structure in the three vertex correlations. In
particular, for scale-free networks it has been observed that in many
instances, the clustering spectrum exhibits also a power-law behavior,
$\barc(k) \sim k^{-\alpha}$. A value of $\alpha$ close to $1$ has been empirically
observed in several natural networks, and analytically found in some
growing network models \cite{ravasz02,szabo,structurednets}. These
findings have led to propose the clustering spectrum $\barc(k)$ as a
tool to measure hierarchical organization and modularity in complex
networks \cite{ravasz02}.

The presence of correlations are thus a very relevant issue in order
to understand and classify complex networks, especially in view of the
important consequences that they can have on dynamical processes
taking place on the topology defined by the networks
\cite{marian1,morenostructured,morenopercolation}. While there are
quite a few empirical results for real networks, the situation is not
the same for network models, and therefore there is no consensus
regarding the origin of assortative and dissasortative mixing, and its
relation with the power law behavior of the clustering spectrum
$\barc(k)$. In fact, most works devoted to analytical calculations of
correlations in complex network models have been performed only for
particular cases \cite{ravasz02,szabo,structurednets,hiddenvars}. In
this respect, it is noteworthy the rate equation formalism proposed by
Szab\'o \textit{et al.} in Ref.~\cite{szabo} (see also \cite{szaboproc})
to compute $\barc(k)$ in growing network models with preferential
attachment. However, and up to our knowledge, no such formalism has
been developed to deal with two vertex correlations, as given by the
$\bark(k)$ function.

In this paper we revise the formalism proposed in Ref.~\cite{szabo}
for computing the clustering spectrum in growing networks models with
preferential attachment. Reconsidering the mean field rate equation in
the continuous degree approximation for the $\barc(k)$ presented in
\cite{szabo}, we are able to provide a general expression for the
boundary condition of this rate equation, which was neglected in the
original treatment and which can have as a matter of fact relevant
effects in the final solution, as we will show below.  Inspired by
this result we also propose a new rate equation for two vertex
correlations, as measured by the $\bark(k)$ function, and work out the
correponding boundary condition.  We remark that both equations are
valid in general for the so-called ``citation networks''
\cite{mendesbook}, in which neither edge or vertex removal nor edge
rewiring are allowed.  The general formalism is presented in section
\ref{sec:rate-equat-corr}. The rate equations obtained can be easily
solved for growing networks with linear preferential attachement (LPA)
\cite{mendes99}, as shown in section ~\ref{sec:lpa}.  In particular,
we are able to obtain expressions for the dependence of the
correlations on the degree $k$ and the system size $N$, for both the
dissasortative and assortative regimes of the model, which are in very
good agreement with numerical simulations and previous scaling
arguments \cite{dorogorev}. LPA models generate networks with a
vanishing average clustering coefficient $C$.  In order to asses the
effects of a nonzero clustering, we study in section
\ref{sec:gener-dorog-mend} a growing model presenting a large final
clustering coefficient \cite{dms}, which we are able to compute with
very good accuracy. The results obtained are qualitatively similar to
those shown by the Holme-Kim model \cite{szabo,holme02c}. The rate
equation proposed for two vertex correlations can be easily
generalized to deal with more involved situations. As an example of
its flexibility, we examine in section \ref{sec:weight-grow-netw} a
recently proposed model for the evolution of weighted complex networks
\cite{barrat04:_weigh}. In this case, we extend our formalism to
compute a function estimating weighted two vertex correlations, in
which the actual strength of the interactions between neighboring
vertices is taken into account.  Our results allow us to discuss the
scaling form of two and three vertex correlation functions, and signal
the possible relations that can be established between them.

\section{Rate equations for correlations in  growing networks}
\label{sec:rate-equat-corr}

Let us consider the class of growing network models in which, at each
time step, a new vertex with $m$ edges is added to the network. For the
vertex introduced at time $t$, each of its emanating edges is
connected to an existing vertex introduced at time $s$ ($s< t)$ with a
connection probability $\Pi_s(\{k\}, t)$, which is assumed to depend only
on the degrees of the existing vertices at time $t$, $\{k\}= \{k_{1}(t), \ldots
k_{t-1}(t)\}$.  Time runs from $1$ to $N$ (the final network size), and
since for each new vertex $m$ edges are added, the average degree is
fixed and given by $\avk=2m$. In the continuous $k$ and $t$
approximation \cite{dorogorev}, the average degree that the vertex $s$
(i.e. the vertex introduced at time $s$) has at time $t$ ($t>s)$ can
be computed from the rate equation
\begin{equation}
  \frac{d k_s(t)}{d t} = m \Pi_s(\{k\}, t),
  \label{eq:7}
\end{equation}
with the boundary condition $k_s(s)=m$ (initially all vertices have
$m$ connections). From $k_s(t)$, the degree distribution can be
obtained as 
\begin{equation}
  P(k, t) = - \frac{1}{t} \left.\left(\frac{\partial k_s(t)}{\partial s}
    \right)^{-1}\right|_{s=s(k,t)} ,
\end{equation}
where $s(k,t)$ is the solution of the implicit equation $k=k_s(t)$. 

For this class of networks it is possible to obtain a rate equation
for the clustering spectrum.  Following Ref.~\cite{szabo}, we recall
that the clustering coefficient $c_s(t)$ of vertex $s$ at time $t$ is
defined as the ratio between the number of edges between the neighbors
of $s$ and its maximum possible value. Then, if $M_s(t)$ is the number
of connections between the neighbors of $s$ at time $t$, we have that
\begin{equation}
  c_s(t) = \frac{2 M_s(t)}{k_s(t) [ k_s(t)-1]}.
  \label{eq:17}
\end{equation}
During the growth of the network, $M_s(t)$ can only increase by the
simultaneous addition of an edge to $s$ and one of its neighbors.
Therefore, in the continuous $k$ approximation, we can write down the
following rate equation \cite{szabo}:
\begin{equation}
  \frac{d M_s(t)}{d t} = m(m-1) \Pi_s(\{k\}, t) 
  \sum_{j \in {\cal V}(s)} \Pi_j(\{k\}, t) \ ,
  \label{eq:3}
\end{equation}
where $\mathcal{V}(s)$ is the set of neighbours of vertex $s$.
In order to solve this equation we must provide additionally a
boundary condition. To do so, we observe that 
 $M_s(s)$ is the
number of triangles created by the introduction of vertex $s$.
Therefore
\begin{equation}
  M_s(s) = \frac{m(m-1)}{2} \sum_{j, n=1}^s \Pi_j(\{k\}, s) \Pi_n(\{k\}, s) \Pi_{j,n},
  \label{eq:4}
\end{equation}
that is, it is proportional to the probability that $s$ is connected
to vertices $j$ and $n$, times the probability $\Pi_{j,n}$ that $j$ and
$n$ are linked, averaged over all vertices $j$ and $n$ existing in the
network at time $s$. The probability $\Pi_{j,n}$ is given by
\begin{equation}
  \Pi_{j,n} = \Theta(j-n) m \Pi_n(\{k\}, j) + \Theta(n-j) m \Pi_j(\{k\}, n),
  \label{eq:5}
\end{equation}
where $\Theta(x)$ is the Heaviside step function. Solving the equation for
$M_s(t)$ with the boundary condition Eq.~(\ref{eq:4}), we can obtain
the clustering $c_s(t)$ from Eq.~(\ref{eq:17}). Then, since in growing
network models in the continuous $k$ approximation the degree at time
$t$ is uniquely determined by the introduction time $s$, from $c_s(t)$
we can directly obtain the clustering spectrum $\barc(k)$ as a
function of $k$ and the largest time $t=N$.

In the case of the two vertex correlation function $\bark(k)$, we can
proceed along similar lines. Let us define $R_s(t)$ as the sum of the
degrees of the neighbors of vertex $s$, evaluated at time $t$. That
is, 
%if we denote by $\mathcal{V}(s)$ the set of neighbors of vertex
%$s$, then
\begin{equation}
  R_s(t) =  \sum_{j \in {\cal V}(s)} k_j(t).
  \label{eq:9}
\end{equation}
The average degree of the neighbors of vertex $s$, $\bark(s)$ is then
given by $\bark(s) = R_s(t) / k_s(t)$.  During the growth of the
network, $R_s(t)$ can only increase by the adjunction of a new vertex
connected either directly to $s$, or to a neighbour of $s$. In the
first case $R_s(t)$ increases by an amount $m$ (the degree of the
newly linked vertex), while in the second case it increases by one
unit. Therefore, in the continuous $k$ approximation, we can write
down the following rate equation:
\begin{equation}
  \frac{dR_s(t)}{dt} = m [m \Pi_s(\{k\}, t)] + m \sum_{j \in {\cal V}(s)}
  \Pi_j(\{k\}, t).
  \label{eq:1}
\end{equation}
In order to obtain the boundary condition for this equation, we observe that,
at time s, the new vertex $s$ connects to an old vertex of degree $k_j(s)$
with probability $\Pi_j(\{k\}, s)$, and that this vertex gains a new
connection in the process. Therefore,
\begin{equation}
  R_s (s) = m \sum_{j=1}^{s} \Pi_j(\{k\}, s) [k_j(s)+1].
  \label{eq:2}
\end{equation}
>From the solution of this rate equation, we can obtain $\bark(s)$ and
from it the two vertex correlation function by the functional
dependence of $s$ on $k$ and $t=N$.

We must note that Eqs.~(\ref{eq:3}), (\ref{eq:4}), (\ref{eq:5}),
(\ref{eq:1}), and~(\ref{eq:2}) are only valid for the so-called
``citation networks'', in which neither edge removal nor rewiring
\cite{albert00} are allowed, since these two processes can induce
nonlocal variations in the conectivity of the nearest neighbors.

\section{Linear preferential attachement models}
\label{sec:lpa}

As an example of the application of the rate equations presented in the
previous Section, we consider the general LPA model proposed in
Ref.~\cite{mendes99}, for which the rate equations for
$R_s(t)$ and $M_s(t)$ can be closed and solved analytically. For
general LPA, the connection probability takes the form
\begin{equation}
  \Pi_s(\{k\}, t) = \frac{b_1 k_s(t)+b_2 }{\sum_j [b_1 k_j(t)+b_2]},
  \label{eq:6}
\end{equation}
where $b_i$ are real constants.  Since, for each new vertex, $m$ edges
are added to the network, the normalization constant in
Eq.~(\ref{eq:6}) takes the form $\sum_j [b_1 k_j(t)+b_2] = (2mb_1+b_2)t$.
Thus, the model depends only on the tuning parameter $a=b_2/b_1$,
taking values in the interval $a \in ]-m, \infty[$ (since the minimum
degree is $m$, $a$ cannot be lower than $-m$ in order for
$\Pi_s(\{k\}, t)$ to remain positive).

Solving the rate equation for the degrees
Eq. (\ref{eq:7}), we obtain
\begin{equation}
k_s(t) = (m+a) \left(\frac{t}{s} \right)^\beta  -a, 
\ \ \beta=\frac{m}{2m+a}. 
\label{eq:ks}
\end{equation}
Therefore, this model yields networks with a power law degree
distribution of the form
\begin{equation}
P(k) \sim  k^{-\gamma}, \ \ \gamma=3+a/m.
\end{equation}
For $a>0$, we obtain a degree exponent $\gamma>3$, which corresponds to
finite degree fluctuations in the thermodynamic limit.  The case $a=0$
recovers the original BA model with $\gamma=3$ \cite{barab99}.  Finaly,
values $-m < a< 0$ yield scale free networks with a tunable degree
exponent, in the range $\gamma\in ]2,3[$.

\subsection{Two vertex degree correlations}

The rate equation for $R_s(t)$ takes in this case the form
\begin{eqnarray}\nonumber
  \frac{dR_s(t)}{dt} &=& m^2 \frac{k_s(t)+a}{(2m+a)t}
  +\sum_{j \in {\cal V}(s)} m \frac{k_j(t)+a}{(2m+a)t} \\
  &=& \beta \frac{(m+a)k_s(t)+am}{t} + \beta 
  \frac{R_s(t)}{t},
  \label{eq:10}
\end{eqnarray}
where we have used the definition of $R_s(t)$, Eq.~(\ref{eq:9}). 
The general solution of the previous equation is 
\begin{equation}
  R_s(t)= A_0(s) t^\beta + 
  \beta (m+a)^2 \left( \frac{t}{s} \right)^\beta \ln t +a^2
  \label{eq:12}
\end{equation}
where $A_0(s)$ is given by the boundary condition $R_s(s)$. From
Eq.~(\ref{eq:2}), we have that 
\begin{eqnarray}
  R_s (s) &=& m \sum_{j=1}^{s} \frac{ a+ (a+1) k_j(s) +
    k_j^2(s)}{(2m+a)s} \nonumber \\
  &=& \beta a + 2 m \beta (a+1) + \frac{\beta}{s}  \sum_{j=1}^{s} k_j^2(s) \ .
\end{eqnarray}
Plugging in 
$k_j (s) = (m+a) (s/j)^{\beta} -a$ into $R_s(s)$ results in
\begin{equation}
  R_s (s) =  m(1-a) + \beta (m+a)^2 s^{2 \beta-1} \sum_{j=1}^{s} j^{-2 \beta}.
  \label{eq:11}
\end{equation}
In order to estimate the behaviour of the previous expression, we have
to distinguish the different cases corresponding to the possible
values of $\beta$ (namely $a$).

\textbf{(i)} $-m < a < 0$ (i.e. $\beta > 1/2$, $\gamma<3$). In this case, for
large $s$, $ \sum_{j=1}^{s} j^{-2 \beta} \simeq \zeta(2\beta)$, where $\zeta(x)$ is the
Riemann Zeta function, and thus, at leading order,
\begin{equation}
  R_s (s) \simeq  \beta \zeta(2\beta) (m+a)^2 s^{2\beta-1}. 
  \label{eq:23}
\end{equation}
The determination of the integration constant $A_0(s)$ from
Eq.~(\ref{eq:12}) yields then
\begin{equation}
  R_s(t) \simeq   \beta \zeta(2\beta) (m+a)^2 t^\beta s^{\beta-1} 
  + \beta (m+a)^2 \left( \frac{t}{s} \right)^\beta 
  \ln \left( \frac{t}{s} \right) ,
\end{equation}
where terms independent of $t$ and $s$ and terms going to zero in the
large $t$ or $s$ limit have been neglected.  From the definition of
$\bark(s)$, and substituting $s$ as a function of $k$ and $t=N$ (the
network final size) in the limit of large $k$ and $N$ we obtain the
following expression for the average degree of the neighbors of the
vertices of degree $k$:
\begin{eqnarray}
  \bark(k,N) &\simeq&  \beta \zeta(2\beta) (m+a)^{3-1/\beta}  N^{2\beta-1}k^{-2+1/\beta}  \nonumber
  \\ &+&(m+a)\ln
  \left( \frac{k}{m+a}\right).
\label{eq:15}
\end{eqnarray}

>From this expression, we conclude that the LPA with $a<0$ yields in
the large $N$ limit networks with dissasortative two vertex
correlations, characterized by a power-law decay $\bark(k,N) \sim
N^{2\beta-1} k^{-2+1/\beta}$. This exponent was previously obtained by scaling
arguments in Ref.~\cite{dorogorev}. The dependence of the prefactor on
$N$ implies that $\bark(k,N)$ diverges in the thermodynamic limit
$N\to\infty$, in agreement with the theoretical arguments provided in
Ref.~\cite{marian3}. For finite $N$, however, the logarithmic term
with constant prefactor can induce corrections to the power-law
scaling.  Since $2\beta -1$ is at most $1$, the growth of $\bark(k,N)$ is
not very steep with $N$ and these corrections are observable in
numerical simulations, as we will see below in this Section.

\textbf{(ii)} $a=0$ (i.e. $\beta = 1/2$, $\gamma=3$). For this value of $\beta$
Eq.~(\ref{eq:11}) is dominated by a logarithmic divergence,
$\sum_{j=1}^{s} j^{-1} \simeq  \ln s$, yielding
\begin{equation}
  R_s(s) \simeq  \frac{m^2}{2}\ln s.
\end{equation}
>From here, we obtain 
\begin{equation}
  R_s(t) \simeq \frac{m^2}{2} \sqrt{\frac{t}{s}} \ln t,
\end{equation}
and finally
\begin{equation}
  \bark(k,N) \simeq  \frac{m}{2} \ln N.
\label{eq:knnBA}
\end{equation}
That is, two vertex correlations are independent of the degree and
grow with the system size as $\ln N$, in agreement with the behavior
expected for an uncorrelated scale-free network with degree exponent
$\gamma=3$, Eq.~(\ref{eq:13}). Numerical simulation of the BA
model \cite{alexei02} show actually a very weak dependence on $k$ in
the $\bark(k,N)$ function, compatible nevertheless with the behavior
given by our rate equation approach in the large $k$ limit. This $k$
dependence, evidentiated at small values of the degree, cannot be
detected within our approach, since it has been formulated in the
continuous $k$ approximation.

\textbf{(iii)} $a>0$ (i.e. $\beta < 1/2$, $\gamma>3$). In this situation, the
summation in Eq.~(\ref{eq:11}), 
$\sum_{j=1}^{s} j^{-2\beta} \simeq s^{1-2 \beta}/(1-2\beta)$, 
is divergent, and therefore $R_s(s)$ becomes independent of $s$.
This leads to
\begin{eqnarray}
  R_s(t) &\simeq& \beta (m+a)^2 \left( \frac{t}{s} \right)^\beta 
  \ln \left( \frac{t}{s} \right)  \nonumber \\
  &+&
  \left[ m(1-a)+ \frac{\beta(m+a)^2}{1-2\beta} -a^2
    \right] \left( \frac{t}{s} \right)^\beta,
\end{eqnarray}
and finally the dominant behaviour for the correlation function is
\begin{equation}
  \bark(k,N) \simeq  (m+a) \ln \left(\frac{k}{m+a} \right) 
\end{equation}
In this case, $\bark(k,N)$ is independent of the network size, and
increases logarithmically with $k$: For $\gamma>3$, LPA yields networks
with weak assortative mixing.

\subsection{Three vertex correlations}

In order to estimate three vertex degree correlations by means of the
clustering spectrum $\barc(k)$, we start from the rate equation
Eq.~(\ref{eq:3}), which for the LPA model takes the form

\begin{eqnarray}\nonumber
  \lefteqn{\frac{d M_s(t)}{d t} = m(m-1) \frac{k_s(t)+a}{(2m+a)t}
  \sum_{j \in {\cal V}(s)} \frac{k_j(t)+a}{(2m+a)t}} \\
  &&= m(m-1) \frac{k_s(t)+a}{(2m+a)^2 t^2} [R_s(t) + a k_s(t)].
  \label{eq:14}
\end{eqnarray}
The boundary condition $M_s(s)$ can be written as

\begin{widetext}
\begin{eqnarray}
  M_s(s) &=& \frac{m(m-1)}{2} \sum_{j, n} \Pi_j(\{k\}, s) \Pi_n(\{k\}, s)
  \Pi_{j, n} \nonumber \\
  &=& \frac{\beta^2 (m-1) (m+a)^3}{2(2m+a)} s^{2 \beta-2} \left\{ \sum_{n=1}^s
      n^{-2 \beta}  \sum_{j=n+1}^s j^{-1} + \sum_{j=1}^s
      j^{-2 \beta}  \sum_{n=j+1}^s n^{-1} \right\} \nonumber \\
    &=& \frac{\beta^2 (m-1) (m+a)^3}{2(2m+a)} s^{2 \beta-2}  
    \times 2 \times \sum_{n=1}^s n^{-2 \beta}  \sum_{j=n+1}^s j^{-1} \ .
    \label{eq:mss}
\end{eqnarray}

In order to solve Eq.~(\ref{eq:14}), we approximate $k_s(t)$ and
$R_s(t)$ by their dominant terms for large $t$ and $s$, as computed
above for the different possible values of $a$.

\textbf{(i)} $-m < a < 0$. In this case we have
\begin{equation}
  k_s(t) \simeq  (m+a) \left(\frac{t}{s} \right)^\beta, \ \ \ \ R_s(t) \simeq
  \beta \zeta(2\beta) (m+a)^2  t^\beta s^{\beta-1}.
\end{equation}
Introducing this expression into Eq.~(\ref{eq:14}), we obtain at
leading order 
\begin{equation}
  M_s(t) \simeq  \beta^2 \frac{(m-1)(m+a)^3 \zeta(2\beta)}{(2 \beta-1)(2m+a)}
  (t^{2\beta-1} - s^{2\beta-1})s^{-1} + M_s(s) .
  \label{eq:18}
\end{equation}
In order to compute $M_s(s)$, we observe that the double summation in
Eq.~(\ref{eq:mss}) takes the form at large $s$
\begin{equation}
  \mathcal{S} =  \sum_{n=1}^s  n^{-2 \beta}  
 \sum_{j=n+1}^s j^{-1} \simeq  \sum_{n=1}^s
  n^{-2 \beta} (\ln s - \ln n)  \simeq  \zeta(2\beta) \ln s,
  \label{eq:19}
\end{equation}
since $\sum_{n=1}^\infty n^{-2 \beta } \ln n$ is convergent for $\beta> 1/2$.
Thus we obtain 
\begin{equation}
  M_s(t) \simeq \beta^2 \frac{(m-1)(m+a)^3}{(2 \beta-1)(2m+a)} (t^{2\beta-1} -
  s^{2\beta-1}) s^{-1} +\beta^2 \frac{(m-1)(m+a)^3}{2m+a} \zeta(2\beta)  
s^{2\beta-2} \ln s,
\end{equation}
and from here the expression for the three vertex correlation
function follows:
\begin{eqnarray}
 \barc(k,N) &\simeq& 
\frac{2 \beta^2 (m-1) (m+a)^{3-1/ \beta}}{(2 \beta-1)(2m+a)}
  N^{2\beta -2}  k^{-2 +1 / \beta} \nonumber  \\
 &+& \frac{2 \beta^2 \zeta(2 \beta)(m-1) (m+a)^{5-2/ \beta}}{2m+a}  \ln N
  N^{2\beta -2}  k^{-4 +2 \beta}.
\label{eq:ckaneg}
\end{eqnarray}

To understand the asymptotic behavior of $\barc(k,N)$, two limits have
to be taken, corresponding to large $N$ and large $k$:
\begin{itemize}
\item At fixed and large $N$, the leading behavior at large $k$ is
  $\barc(k,N) \sim N^{2\beta -2} k^{-2 +1 / \beta}$.
\item At fixed $k \lesssim (\ln  N)^{\beta/(2\beta-1)}$ and large $N$, 
the leading
  behavior is instead $\barc(k,N) \sim N^{2\beta -2} \ln N k^{-4 +2 / \beta}$.
\end{itemize}
Therefore, in the numerical simulations we should expect to observe a
crossover between these two scaling regimes.

\textbf{(ii)} $a=0$. We now have
\begin{equation}
  k_s(t) \simeq  m \sqrt{\frac{t}{s}}, \ \ \ \ R_s(t) \simeq
  \frac{m^2}{2} \sqrt{\frac{t}{s}} \ln t,
\end{equation}
which yields
\begin{equation}
  M_s(t)\simeq  \frac{m^2(m-1)}{16 s} ( \ln^2 t - \ln^2 s ) + M_s(s).
\end{equation}
Since $\beta=1/2$, $M_s(s)$, as given by Eq. (\ref{eq:mss}), can be easily
shown to be
\begin{equation}
M_s(s) = \frac{m^2(m-1)}{16 s} \ln^2 s \ ,
\label{eq:bcBA}
\end{equation}
and we obtain
\begin{equation}
  M_s(t)\simeq  \frac{m^2(m-1)}{16 s} \ln^2 t ,
\end{equation}
which results in a clustering coefficient at large $N$
\begin{equation}
  \barc(k,N)\simeq  \frac{m-1}{8} \frac{\ln^2 N}{N}.
\end{equation}

We recover the well-known result for the BA model that the clustering
spectrum is constant, and scaling as $\ln^2 N/N$, as observed in
Ref.~\cite{szabo}. It is worth noting that the computation
of the boundary condition (\ref{eq:bcBA}) is essential in recovering this 
result.

\textbf{(iii)} $a>0$. For this range of values of $a$ we have
\begin{equation}
  k_s(t) \simeq  (m+a) \left(\frac{t}{s} \right)^\beta, \ \ \ \ R_s(t) \simeq   \beta
  (m+a)^2 \left( \frac{t}{s} \right)^\beta  
  \ln \left( \frac{t}{s} \right),
\end{equation}
yielding
\begin{equation}
  M_s(t) \simeq  \beta^2 \frac{(m-1)(m+a)^3}{(2m+a)(1-2\beta)} s^{-2 \beta}
  \left\{ -t^{2\beta-1} \ln \left(\frac{t}{s}\right) + \frac{ s^{2\beta-1} -
      t^{2\beta-1}}{1-2 \beta} \right\}+ M_s(s).
\end{equation}
For the evaluation of $M_s(s)$, we observe that the double summation
$\mathcal{S}$ defined in Eq.~(\ref{eq:19}) shows now a power-law
divergence, $\mathcal{S} \simeq s^{1-2\beta} /(1-2\beta)^2$. Thus we have
\begin{equation}
  M_s(t) \simeq \beta^2 \frac{(m-1)(m+a)^3}{(2m+a)(1-2\beta)} s^{-2\beta} \left\{
    -t^{2\beta-1}  \ln \left(\frac{t}{s}\right) + \frac{2 s^{2\beta-1} 
      - t^{2\beta-1}}{1-2\beta} \right\},
\end{equation}
yielding a three vertex correlation function
\begin{equation}
  \barc(k,N) \simeq  
  \frac{4 \beta^2(m-1) (m+a)^{3-1/ \beta}}{(2m+a)(1-2\beta)^2} N^{-1}
  k^{-2+ 1/ \beta}. 
  \label{eq:20}
\end{equation}
Therefore, for $a>0$ (i.e. $\beta < 1/2$), 
we obtain that the average clustering of the
vertices of degree $k$ is a growing function of $k$, scaling as
$\barc(k,N) \sim N^{-1} k^{-2+ 1/ \beta}$. Since by definition the clustering
must be smaller than $1$, this growing behavior must be restricted to
degree values $k \lesssim N^{\beta/(1-2\beta)}$.
\end{widetext}

\subsection{Computer simulations}

In order to check the analytical results obtained in this Section, we
have performed extensive numerical simulations of the LPA model.
Simulations were performed for system sizes ranging from $N=10^3$ to
$N=10^6$, averaging over $100$ network samples for each value of $N$
and $a$. We focus in particular in the ranges $a<0$ and $a>0$, which
have not been previously explored (for numerical data corresponding to
$a=0$, the BA model, see Refs.~\cite{alexei02,klemm02b}).

\begin{figure*}
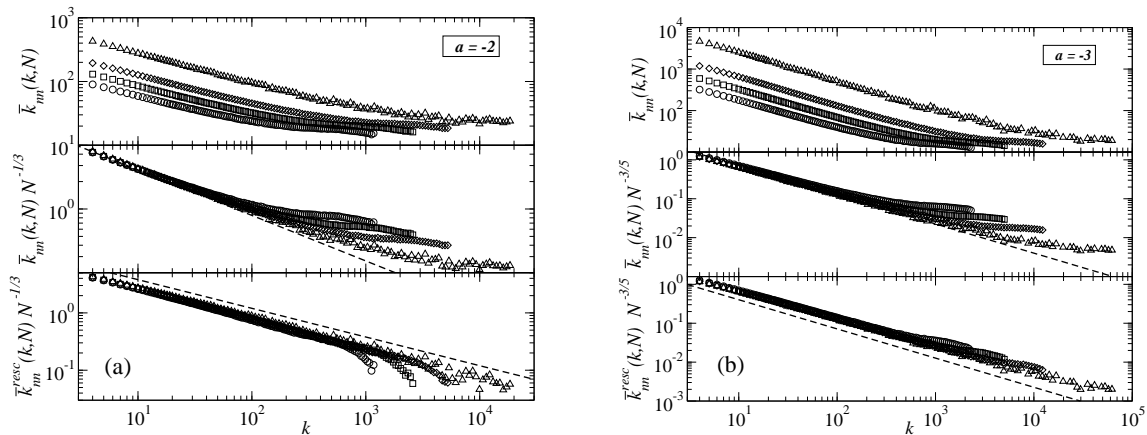

  \centerline{
    \epsfig{file=fig1a.eps, width=7cm}\hspace*{1cm}
    \epsfig{file=fig1b.eps, width=7cm}
  }
  \caption{Average degree for the nearest neighbors of the vertices of
    degree $k$, $\bark(k,N)$, for the LPA model for $m=4$, with $a=-2$
    (a) and $a=-3$ (b).  Symbols correspond to the different system
    sizes $N=10^4$ ($\circ$), $N=3 \times 10^4$ ($\Box$) $N=10^5$ ($\diamond$) $N=10^6$
    ($\triangle$).  Top plots: Raw data.  Middle plots: Data rescaled by the
    size prefactor $N^{1-2\beta}$.  Bottom plots: Data rescaled by the
    size prefactor with logarithmic corrections.  The dashed lines
    represent a power-law decay with exponent $-2+1/ \beta$.}
\label{fig:knnlin}
\end{figure*}

\begin{figure*}
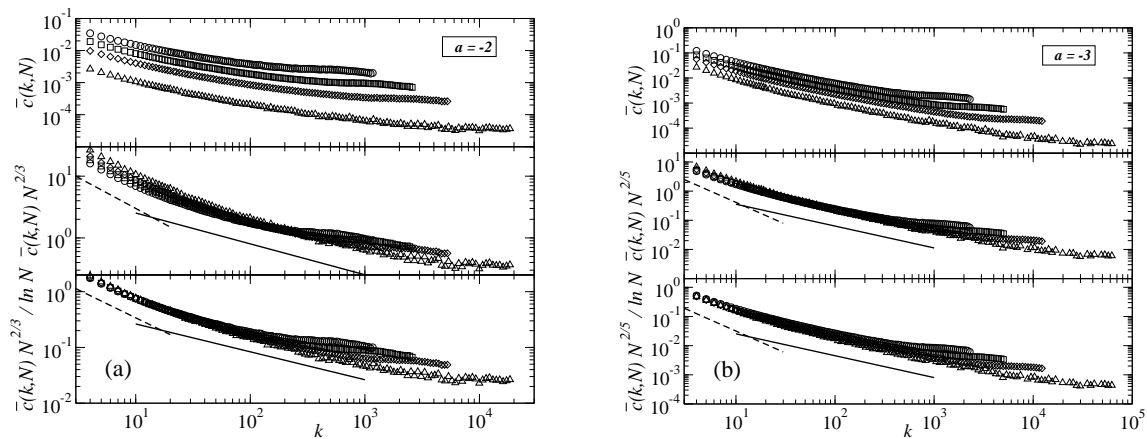

  \centerline{
    \epsfig{file=fig2a.eps,  width=7cm}\hspace*{1cm}
    \epsfig{file=fig2b.eps, width=7cm}
  }
\caption{Clustering spectrum $\barc(k,N)$ for the LPA  model for
  $m=4$, with $a=-2$ (a) and $a=-3$ (b).  Symbols correspond to the different
  system sizes $N=10^4$ ($\circ$), $N=3 \times 10^4$ ($\Box$) $N=10^5$
  ($\diamond$) $N=10^6$ ($\triangle$).  Top plots: Raw data.  Middle plots:
  Data rescaled by the size prefactor $N^{2-2 \beta}$, corresponding to large
  $k$.  Bottom plots: Data rescaled by the size prefactor$N^{2-2
    \beta}/\ln(N)$, corresponding to small $k$.  The full and dashed lines
  represent power-law decays with exponent $-2+1/ \beta$ and $-4+2/\beta$,
  respectively.}
\label{fig:ckaneglin}
\end{figure*}

In Figs.~\ref{fig:knnlin} and~\ref{fig:ckaneglin} we explore the
behavior of networks generated for $a<0$. We consider first the
average degree of the nearest neighbors $\bark(k,N)$.
Fig.~\ref{fig:knnlin}(a) corresponds to $m=4$, $a=-2$, values that
yield $\beta=2/3$ and $\gamma=2.5$, while Fig.~\ref{fig:knnlin}(b) plots data
for $m=4$, $a=-3$, corresponding to $\beta=4/5$ and $\gamma=2.25$. The dashed
lines represent the power law behavior $k^{-2+1/ \beta }$ expected
analytically.  We observe that, as the size of the network increases,
the data follow the predicted scaling $\bark(k,N) \sim N^{2\beta-1} k^{-2+1/
  \beta}$ on larger and larger ranges.  Nevertheless, the logarithmic
corrections present in Eq.~(\ref{eq:15}) are clearly visible from the
large $k$ deviations shown by the data (middle plots in
Fig.~\ref{fig:knnlin}).  The logarithmic correction can, in fact, be
taken into account if one rescales $\bark(k,N)$ appropriately. Namely,
if we define
\begin{equation}
  \bark^\mathrm{resc}(k,N) =  \bark(k,N) - (m+a)\ln \left(
    \frac{k}{m+a}\right), 
\end{equation}
then, from see Eq.~(\ref{eq:15}), we expect
\begin{equation}
  \bark^\mathrm{resc}(k,N)  N^{1-2\beta} \sim k^{-2+1/\beta}.
\end{equation}
In the bottom plots of Fig.~\ref{fig:knnlin} we draw the rescaled
average degree of the nearest neighbors with logarithmic corrections. The
collapse of the data is indeed surprisingly good, given the numerous
approximations and leading order cancellations made in our
calculations. The remaining discrepancy at very large $k$ is
presumably due to the subdominant terms we have neglected.

In Fig.~\ref{fig:ckaneglin} we represent the clustering spectrum
$\barc(k,N)$ for the same parameters as before, i.e. $m=4$, $a=-2$ (a)
and $m=4$, $a=-3$ (b). The top plots represent the corresponding
nonrescaled raw data.  According to the solution provided in
Eq.~(\ref{eq:ckaneg}), for small values of $k$ it is expected an
asymptotic scaling of the form $\barc(k,N) \sim N^{2\beta-2} \ln N k^{-4+2/
  \beta}$. This behavior is well recovered in the bottom plots in
Fig.~\ref{fig:ckaneglin} for both values of $a$, where we can see that
the first points in the graphics for different values of $N$ collapse
onto the same curve, with the predicted $k$ dependence. For large
values of $k$, on the other hand, we expect instead a scaling
$\barc(k,N) \sim N^{2\beta-2} k^{-2+1/ \beta}$, which is again recovered in the
middle plots of this Figure, showing a better collapse in the
intermediate range of $k$ values.  At very large $k$ values, finally,
the neglected logarithmic terms come into play, affecting the scaling
of the data. It is important to notice the important role played by the
boundary condition Eq.~(\ref{eq:4}), which is responsible for the
second term in Eq.~(\ref{eq:ckaneg}), giving the correct scaling behavior
for small $k$.

\begin{figure*}
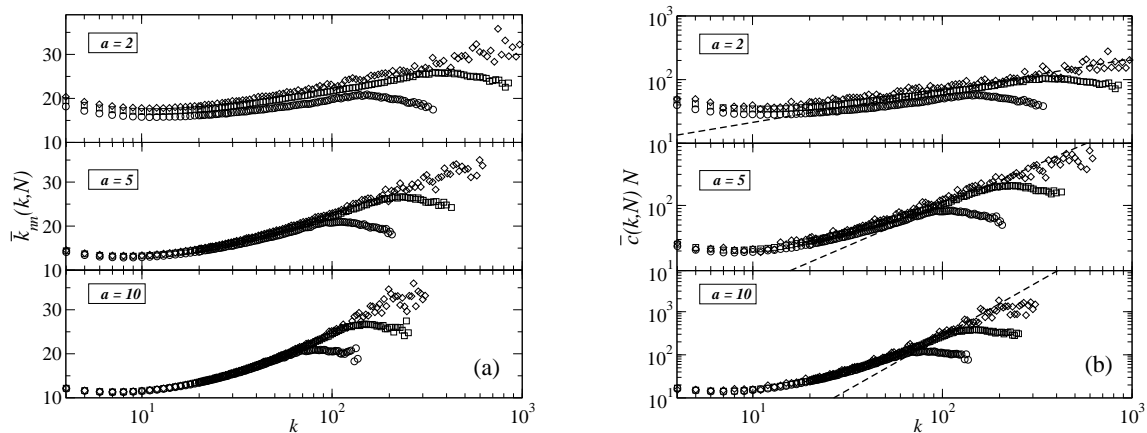

\centerline{
  \epsfig{file=fig3a.eps, width=7cm}\hspace*{1cm}
  \epsfig{file=fig3b.eps, width=7cm}
}
\caption{Average degree for the nearest neighbors of the vertices of
  degree $k$, $\bark(k,N)$ (a) and clustering spectrum, $\barc(k,N)$
  (b) for the LPA with $m=4$ and positive values of $a$.  Symbols
  correspond to the different system sizes $N=10^4$ ($\circ$), $N=10^5$
  ($\Box$) $N=10^6$ ($\diamond$).  Top plots: $a=2$.  Middle plots: $a=5$.
  Bottom plots: $a=10$. Data for $\barc(k,N)$ has been rescaled by the
  theoretical size prefactor $N$. The dashed lines represent a
  power-law behavior with exponent $-2+1/ \beta$.}
\label{fig:positivelin}
\end{figure*}

In Fig.~\ref{fig:positivelin} we finally explore the average degree of
the nearest neighbors (a) and the clustering spectrum (b) for the LPA
model with $a>0$. We focus in particular on the values $m=4$ and $a=2$
(top plots), yielding $\beta=2/5$, $\gamma=3.5$; $a=5$ (middle plots), with
$\beta=4/13$, $\gamma=4.25$; and $a=10$ (bottom plots), that corresponds to
$\beta=2/9$, $\gamma=5.5$.  For the $\bark(k,N)$ function our theoretical
analysis predicts a function independent of the network size, and
slowly (logarithmically) growing with the degree.  These predictions
are confirmed in Fig.~\ref{fig:positivelin}(a).  It is noteworthy that
the theoretical prediction becomes more accurate for large $a$: While
the collapse is quite good for $a\geq5$, the graphs are a bit scattered
for the smallest value of $a$ considered. This fact is due to the slow
convergence (as $N$ grows) to the theoretical asymptotic form for
small $a$. Analogously, the clustering spectrum shows the predicted
scaling $\barc(k,N) \sim N^{-1} k^{-2+1/ \beta}$,
Fig.~\ref{fig:positivelin}(b). The dependence on system size is
correctly captured by our analysis for larger values of $a$.  In this
range, however, the power-law dependence on $k$ seems to depart from
the theoretical exponent $-2+1/ \beta$. This apparent departure can be due
to the limited range of degrees for such large values of the degree
exponent (the degree range decreases for increasing $a$), as well as
to the subdominant terms neglected in the asymptotic expression
Eq.~(\ref{eq:20}).

\section{Growing networks with large clustering}
\label{sec:gener-dorog-mend}

As we have seen in the previous Section, the LPA model yields a
clustering spectrum $\barc(k)$ that, even if presenting a non-trivial
scaling, vanishes in the thermodynamic limit, i.e., $\lim_{N\to\infty}
\barc(k,N) = 0$.  However, for many complex networks, such as the
Internet \cite{romuvespibook}, we observe a function $\barc(k)$
scaling with $k$, together with a finite clustering.

Several models have been proposed which reproduce this feature.  In
particular, Dorogovtsev, Mendes and Samukhin (DMS) introduced in
Ref.~\cite{dms} a scale-free growing network with large clustering
coefficient $C$. The model is defined as follows: At each time-step, a
vertex is added and connected to {\em the two extremities of a
  randomly chosen edge}, thus forming a triangle. The resulting
network has a power-law degree distribution $P(k) \sim k^{-3}$, with
$\avk=4$, and since each new vertex induces the creation of at least
one triangle, we expect this model to generate networks with finite
clustering coefficient.  We consider here a generalization of the DMS
model, in which every new node is connected to the extremities of
$m/2$ randomly chosen {\em edges}, where $m$ is an even number. The
original model corresponds thus to $m=2$, and this generalization
allows to tune the average degree, setting it to $\avk=2m$.

It is important to notice that this model actually contains the LPA
mechanism in a disguised form. Indeed, the probability to choose a
vertex $s$ is clearly proportional to the number of edges arriving to
$s$, i.e. to its degree $k_s$. At time $t$ there are $mt$ edges so
that $\sum_s k_s=2mt$ and the probability to choose $s$ when choosing one
edge is $k_s/(mt)$ ($\sum_s k_s/(mt)=2$ since one chooses indeed $2$
vertices). This process is repeated $m/2$ times and thus, at each time
step the probability to choose $s$ is $k_s/(2t)$.

This shows that another way of formulating the random choice of an
edge is in fact the following: a vertex $s$ is chosen with the usual
preferential attachment probability $k_s/(2mt)$, and then one of its
neighbors is chosen at random, i.e. with probability $1/k_s$.

It is then clear that the rate equation for the degree is given by
\begin{equation}
\frac{d k_s(t)}{d t} = \frac{k_s(t)}{2t},
\end{equation}
leading to $k_s(t) = m (t/s)^{1/2}$ and to a scale free degree
distribution of the form $P(k) \approx 2 m^2 k^{-3}$.

We are now in position to write down the rate equations for the
network correlations, taking into account that, each time a vertex is
chosen, so is one of its neighbours.

\subsection{Two vertex degree correlations}

At each time step, $R_s(t)$ can increase either if the vertex $s$ is
chosen (and then $R_s$ increases by $m+1$ because a neighbour of $s$
is also chosen), or if a neighbour $j$ is chosen together with a
neighbour $l$ of $j$, with $l \neq s$ (and then $R_s$ increases by
$1$). Therefore, we have that
\begin{eqnarray}
\frac{dR_s(t)}{dt} &=& (m+1) \frac{k_s(t)}{2t} +
\sum_{j \in {\cal V}(s)} \frac{k_j(t)}{2t} \left( 1 - \frac{1}{k_j(t)}
\right)\nonumber \\
&=&\frac{m k_s(t)}{2t} + \frac{R_s(t)}{2t} \ .
\end{eqnarray}
This is exactly the same equation than for the LPA with $a=0$, i.e.
the BA model.  Moreover, the boundary condition for $R_s(s)$ can be
written as
\begin{equation}
R_s(s) = \frac{m}{2} 
\sum_{j=1}^s \frac{k_j(s)}{2ms} \left\{
k_j(s) + 1 + \sum_{l \in {\cal V}(j)} \frac{1}{k_j(s)} [k_l(s) +1]
\right\} \ ,
\end{equation}
where, for each one of the $m/2$ edges chosen by $s$, the first term
corresponds to the contribution of $j$ (chosen with probability
$k_j/(2ms)$, and the second term to the contribution of a neighbour
$l$ of $j$, which is chosen with probability $1/k_j$.  This expression
is easily reduced to
\begin{equation}
R_s(s) = 
\frac{1}{2} \sum_{j=1}^s \frac{k_j(s)[k_j(s)+1]}{s} \simeq \frac{m^2}{2} \ln s.
\end{equation}
Once again we obtain the same result than for the LPA with $a=0$. The
conclusion is that the $\bark(k,N)$ function for the generalized DMS
model is given by Eq.~(\ref{eq:knnBA}): the two vertex correlations
are independent of the degree and growing with the network size as
$\ln N$, in the same fashion as the BA model.

\subsection{Three vertex correlations}

In order to write down the rate equation for $M_s(t)$, we have to take
into account that, at each time step, $m/2$ triangles are formed by
the choice of $m/2$ edges, and that, moreover, additional triangles
may be formed for $m>2$ by choosing two different edges with a common
vertex. At each time step, the increase in $M_s(t)$ is thus given by
two terms:

\begin{itemize}
\item The first one comes from choosing the vertex $s$ with
  probability $k_s(t)/(2t)$. In this case, $M_s$ increases by
  $1$,  since one of the neighbours of $s$ is also chosen. 
\item The second contribution comes from the following situation: one
  of the edges chosen is $s-l$, and another one is $j-l'$, with $j \in
  {\cal V}(s)$, $j \neq l$ and $l' \neq s$.
\end{itemize}

The resulting rate equation reads:
\begin{eqnarray}
\frac{dM_s(t)}{dt} &=& \frac{k_s(t)}{2t} + 
\frac{m}{2} \left( \frac{m}{2} -1 \right) \frac{k_s(t)}{mt} \nonumber
\\
&\times& \sum_{j \in {\cal V}(s)}
\frac{k_j(t)}{mt}\left(1-\frac{1}{k_s(t)}\right) \nonumber \\
 &=& \frac{k_s(t)}{2t} + \frac{m-2}{4mt^2} [k_s(t) - 1] R_s(t).
\end{eqnarray}
We use $k_s \simeq  m \sqrt{t/s}$ and $R_s \simeq  m^2 \ln t \sqrt{t/s}/2$
to obtain
\begin{equation}
\frac{dM_s(t)}{dt} \simeq  \frac{m}{2\sqrt{t s}} + \frac{m^2(m-2)\ln t}{8 s
  t}, 
\end{equation}
whose solution reads
\begin{equation}
M_s(t) \simeq  m\left(\sqrt{\frac{t}{s}} -1\right)
+ \frac{m^2(m-2)}{16 s} (\ln^2 t - \ln^2 s) +M_s(s).
\end{equation}

The boundary condition is again given by two contributions: First,
$m/2$ triangles are created by attaching $s$ to $m/2$ edges. The
second contribution is given by
\begin{equation}
\frac{m}{2} \left( \frac{m}{2} -1 \right)
\sum_{j=1}^s 
\sum_{l \in {\cal V}(j)}
\frac{k_j(s)}{ms} 
\left( 1 - \frac{1}{k_j(s)} \right)
\frac{k_{l}(s)}{ms}
\end{equation}
i.e. the sum over all vertices $j$ of the probability that, among the
$m/2$ edges chosen by the new node $s$, one has $j$ for extremity, and
another one has a neighbour $l$ of $j$ for extremity. This yields
\begin{eqnarray}
M_s(s) &=& \frac{m}{2} + \frac{m-2}{4 m s^2} \sum_{j=1}^s R_j(s) [k_j(s)
-1] \nonumber \\
&\simeq&   \frac{m}{2} + \frac{m^2(m-2)\ln^2 s}{8s} 
\end{eqnarray}
and finally
\begin{equation}
M_s(t) \simeq  m \sqrt{\frac{t}{s}} - \frac{m}{2}
+  \frac{m^2(m-2)}{16 s} (\ln^2 t + \ln^2 s) \ .
\end{equation}
The clustering spectrum  can therefore be written as
\begin{equation}
\barc(k,N) \simeq  \frac{2k-m}{k(k-1)}
+\frac{m-2}{8N} 
\left( \ln^2 N + \ln^2 \left(\frac{m^2 N}{k^2}\right) \right) \ .
\label{eq:ckdms}
\end{equation}

The clustering spectrum is now finite in the infinite size limit,
\begin{equation}
  \barc(k) = \lim_{N\to\infty} \barc(k,N) \simeq \frac{2k-m}{k(k-1)}.
\end{equation}
It is interesting to see that, for the original model with $m=2$, the
finite-size corrections actually vanish and we obtain the result
$\barc(k,N) = 2/k$, independent of $N$. This scaling is also similar
to that obtained for the Holme-Kim model in \cite{szabo}. The
knowledge of the exact form of the degree distribution for $m=2$,
$P(k)=12/(k(k+1)(k+2))$ \cite{dms}, allows us to obtain the average
clustering coefficient $C(m=2)=2\pi^2 -19 \approx 0.739$.  More generally, for
large $m$, approximating $P(k)$ by $2m^2/k^3$, and sums by
integrals, yields
\begin{eqnarray}
\lefteqn{C(m) = \int_m^\infty  P(k) \barc(k) \; dk}   \nonumber \\
&&\simeq  2m^2 -3m -4/3 +
2m^2(2-m)\ln\left(\frac{m}{m-1}\right). \quad \quad
\label{eq:cdms}
\end{eqnarray}

\subsection{Computer simulations}

\begin{figure}
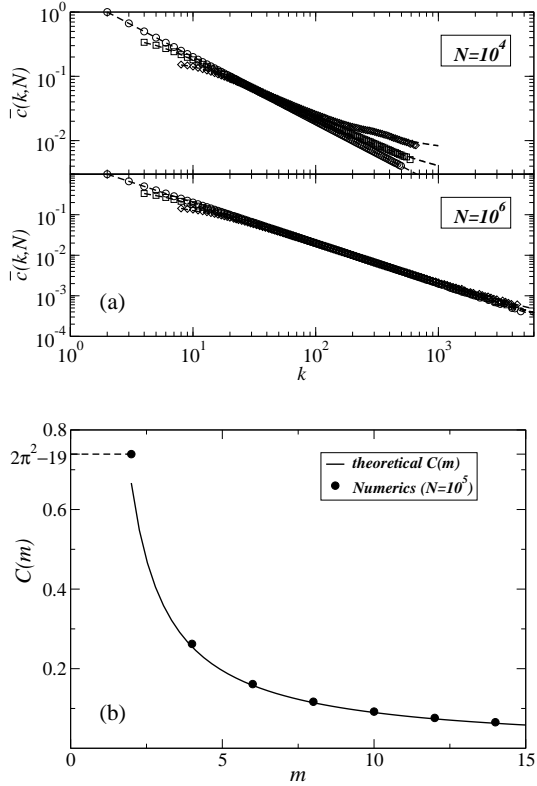

  \centerline{\epsfig{file=fig5a.eps,  width=7cm}}\vspace*{0.5cm}
  \centerline{\epsfig{file=fig5b.eps,  width=7cm}}
\caption{(a) Clustering spectrum $\barc(k,N)$ for the
  generalized DMS model. The top plot corresponds to a systems size
  $N=10^4$; the bottom plot is for $N=10^6$. Symbols correspond to
  different values of the average degree: $m=2$ ($\circ$), $m=4$
  ($\Box$), $m=8$ ($\diamond$).  The dashed lines are the theoretical
  predictions given by Eq.~(\ref{eq:ckdms}). (b) Average clustering
  coefficient $C(m)$ as a function of $m$, for networks of size
  $N=10^5$. The full line represents the theoretical prediction
  Eq.~(\ref{eq:cdms}). The dashed line marks the theoretical value for
  $m=2$, $C(2)=2\pi^2 -19$.}
\label{fig:ckdms}
\end{figure}

We have performed extensive numerical simulations of the generalized
DMS model studied in this Section. We focus on the clustering spectrum
$\barc(k,N)$ since the results for $\bark(k)$ are expected to be equal
to the case of the BA model. Fig.~\ref{fig:ckdms}(a) shows the
excellent agreement between the predicted behaviour,
Eq.~(\ref{eq:ckdms}), and the numerical data for various values of $m$
and sizes ranging from $N=10^4$ to $N=10^6$. As expected, no finite
size corrections are present for $m=2$, while they are correctly
described by the analytical approach for larger $m$.  Moreover, the
prediction for the average clustering coefficient $C(m)$,
Eq.~(\ref{eq:cdms}), is also shown to be in excellent agreement with
numerical data, Fig.~\ref{fig:ckdms}(b).

\section{Weighted growing networks}
\label{sec:weight-grow-netw}

In the previous Sections we have applied the rate equation formalism
to analyze two and three vertex correlations in standard models with
either vanishing or constant clustering coefficient. The formalism for the
two vertex correlations, however, is not limited to these particular
cases, and can be easily extended to analyze more complex growing
network models. As an example, in this Section we will consider a
recently proposed growing weighted network model
\cite{barrat04:_weigh}.  Weighted networks \cite{wiegtedgeneral} are a
natural generalization of graphs in which a real quantity is assigned
to each edge, representing the importance or weight $w_{ij}$ of the
interaction between the vertices $i$ and $j$.  Recently
\cite{barratairport}, it has been pointed out that real weighted
networks present a complex architecture, characterized by broad
distributions of weights, as well as nontrivial correlations between
the values of the weights and the topological structure of the
network.

Motivated by these findings, Ref.~\cite{barrat04:_weigh} proposed a
dynamic growing weighted network model, in which new edges are
attached to old vertices with a connection probability depending on
the strength, or total weight, of the vertex. In order to define the
model, let us consider a weighted network characterized by the
elements $w_{ij}$ defining the weight assigned to the edge connecting
vertices $i$ and $j$. We assume the elements $w_{ij}$ to be symmetric,
that is, $w_{ij}=w_{ji}$. Each vertex $i$ is characterized by both its
degree $k_i$ and its strength $\sigma_i$, defined as
\begin{equation}
  \sigma_i=\sum_{j \in \mathcal{V}(i)}w_{ij}.
\end{equation}
For non-weighted networks, in which $w_{ij} = \delta_{ij}$, we obviously
recover $\sigma_i = k_i$.  The model proposed in
Ref.~\cite{barrat04:_weigh} considers a growing network in which at
each time step, a new vertex is added to the system and connected with
$m$ edges to older vertices. The probability that the new vertex $t$
is connected to $s$ ($s<t$) is given by the connection probability
\begin{equation}
  \Pi_s(\{s\}, t) = \frac{\sigma_s(t)}{\sum_j \sigma_j(t)},
  \label{eq:16}
\end{equation}
that is, linearly proportional to the strength of the old vertex
$s$. Each new edge carries an initial weight $w_0=1$. Additionally,
there is a dynamic rearrangement of the weights belonging to the edges
of the receiving vertex: When the vertex $s$ receives a new
connection, the weight of its edges is increased by an amount
\begin{equation}
  w_{s j}\to w_{s j} + \delta\frac{w_{s j}}{\sigma_s}, 
\quad j \in \mathcal{V}(s).
  \label{rule}
\end{equation}
This rule implies that, for each new vertex added, the total strength
of the network is increased by an amount $2 m + 2 m\delta$, therefore the
normalization constant in Eq.~(\ref{eq:16}) is $\sum_j \sigma_j(t) =
2m(1+\delta)t$. It can be shown, within the continuous $k$
approximation \cite{barrat04:_weigh}, that this model generates
scale-free networks, characterized by the quantities 
\begin{equation}
  \sigma_s(t)=m ~\left(\frac{t}{s}\right)^\beta, 
\quad k_s(t)=\frac{\sigma_s(t) +
    2m\delta}{2\delta+1}, \quad P(k) \sim k^{-\gamma}, \label{eq_ss.vs.t}
\end{equation}
with exponents
\begin{equation}
\beta=\frac{2\delta+1}{2\delta+2}, \quad  \gamma=\frac{4\delta+3}{2\delta+1}.
\end{equation}
Therefore, for $\delta>0$, this model yields power-law degree
distributions with degree exponent $\gamma \in ]2, 3[$ and $\beta>1/2$. The
case $\delta=0$ recovers the BA model.

It is easy to see that, at the level of the mean field rate equations
in the continuous $k$ approximation, the weighted growing network
model described above can be mapped into a growing network with LPA
and negative parameter $a$ given by
\begin{equation}
  a= - \frac{2 m \delta}{2 \delta+1}.
  \label{eq:24}
\end{equation}
\begin{widetext}
Therefore, we expect to observe the two and three vertex correlation
functions
\begin{eqnarray}
  \bark(k,N) &\simeq&  \frac{m \zeta(2\beta)}{2(1+\delta)}
  \left( \frac{m}{2\delta+1} \right)^{2-1/ \beta }
  N^{2 \beta - 1}  k^{-2 +1/ \beta } 
 +\frac{m}{2\delta+1}\ln \left(\frac{2 \delta+1}{m} k \right), \label{eq:21}\\ 
 \nonumber \barc(k,N) &\simeq & \frac{(m-1)(2\delta+1)^2}{4\delta(\delta+1)^2}
  \left(\frac{m}{2\delta+1}\right)^{2-1/   \beta}  
N^{2 \beta - 2}  k^{-2 +1/ \beta } \\
  &+& \zeta(2\beta)\frac{(m-1)(2\delta+1)^2}{4(\delta+1)^3} 
 \left( \frac{m}{2\delta+1} \right)^{4-2/ \beta }
\ln N N^{2 \beta - 2}  k^{-4 +1/ \beta}.
  \label{eq:22}
\end{eqnarray}

\end{widetext}

\subsection{Weighted two vertex correlations}

The definition of the $\bark(k)$ function we have computed above
completely neglects the effect of the weights. Therefore, it provides
a biased view of real correlations in the system (for example, two
neighbors with the same degree but widely different weights give the
same contribution). In order to take into account the effect of the
weights associated to the edges, it has been proposed a new
correlation measure, the weighted average degree of the nearest
neighbors, $\bark^w(k)$ \cite{barratairport},
% This quantity is 
defined as follows:
%For the vertex $s$ at time time $t$, we define the
%weighted average degree of its nearest neighbors as
\begin{equation}
  \bark^w(s) = \frac{1}{\sigma_s(t)} 
\sum_{j \in \mathcal{V}(s)} w_{s j}(t) k_j(t).
\end{equation}
This definition implies that $\bark^w(s) > \bark(s)$ if the edges with
largest weight point to the neighbors with largest degree, while
$\bark^w(s) < \bark(s)$ in the opposite case. Therefore, $\bark^w(s)$
measures the effective affinity to connect with large or small degee
neighbors, according to the magnitude of the interaction weight.  The
weighted average degree of the nearest neighbors $\bark^w(k)$, is
defined as the average of $\bark^w(s)$ for all the vertices with the
same degree $k$.

We can study analytically the
weighted two vertex correlations by seeking a rate equation for the
quantity 
\begin{equation}
  Q_s(t) = \sum_{j \in {\cal V}(s)} w_{s j}(t) k_j(t)
\end{equation}
According to the rules defining the model, at each time step $Q_s(t)$
can increase its value by two mechanisms:
\begin{itemize}
\item If a new vertex is directly attached to $s$, $Q_s(t)$ increases
  by an amount $m+\delta Q_s/\sigma_s$
\item If a new vertex is attached to a neighbour $j$ of $s$, then $Q_s(t)$
  increases by $w_{s j}+\delta w_{s j}/\sigma_j + \delta w_{s j} k_j/\sigma_j$
\end{itemize}
Therefore,  the rate equation fulfilled by $Q_s(t)$ is
\begin{eqnarray}
\lefteqn{\frac{dQ_s(t)}{d  t} = m \Pi_s(\{\sigma\}, t) 
\left( m+ \frac{\delta}{\sigma_s(t)}Q_s(t) \right)} \nonumber \\
&+& \sum_{j \in {\cal V}(s)} m \Pi_j(\{\sigma\}, t) 
\left( w_{s j}+\delta \frac{w_{s j}}{\sigma_j(t)} + 
\delta w_{s j} \frac{k_j(t)}{\sigma_j(t)} \right) \quad \quad
\end{eqnarray}
which, in terms of $\sigma_s(t)$ and $Q_s(t)$, yields
\begin{equation}
\frac{dQ_s(t)}{dt} =
\left(\beta + \frac{\delta}{1+\delta}\right) \frac{Q_s(t)}{t}
+\frac{m+\delta-2m\delta}{2(1+\delta)}\frac{\sigma_s(t)}{t}.
\end{equation}
Inserting the value of $\sigma_s(t)$ given by Eq.~(\ref{eq_ss.vs.t}), the
general solution of this equation is
\begin{equation}
Q_s(t)= A_0(s) t^{\beta+\delta/(1+\delta)} 
- \frac{m}{2\delta} (m+\delta-2m\delta) \left(\frac{t}{s}\right)^\beta.
\end{equation}
Since all new edges have an initial weight $w_0=1$, the initial
condition for $Q_s(t)$ coincides with that of $R_s(t)$. Solving for
$A_0(s)$ from Eq.~(\ref{eq:23}), substituting for the corresponding
value of $a$ given by Eq.~(\ref{eq:24}), we finally obtain in the
large $k$ and $N$ limit
\begin{equation}
  \bark^w(k,N) \simeq 
  \frac{m\zeta(2\beta)}{2(1+\delta)(2 \delta +1)} N^{2 \beta -1}.
  \label{eq:25}
\end{equation}
That is, in this model the weighted average degree of the nearest
neighbours is independent of $k$, signaling the absence of two vertex
weighted correlations, as indeed found numerically in
Ref.~\cite{longweighted}.  There is, however, a scaling with the
system size, given by the factor $N^{2 \beta -1}$, which is the same as
that found for the nonweighted correlations for the same value of $\gamma$.

\subsection{Computer simulations}

\begin{figure}
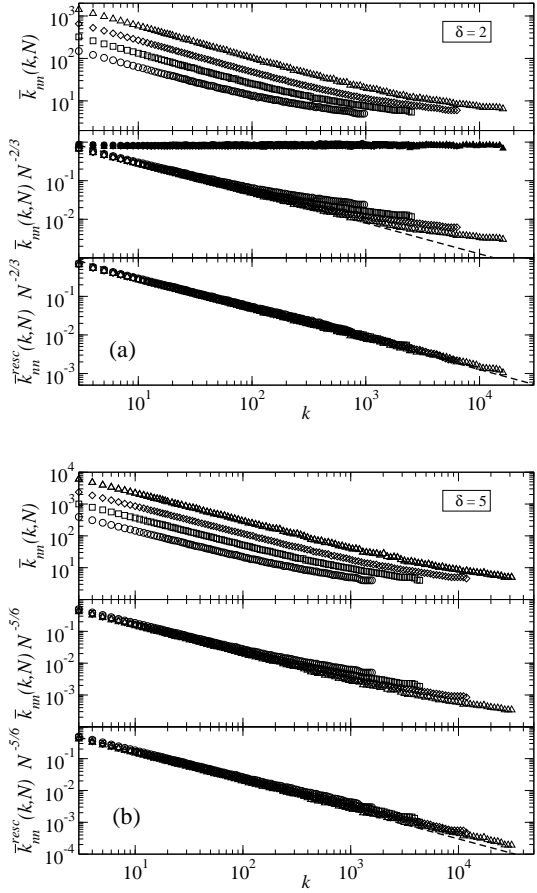

  \centerline{\epsfig{file=fig4a.eps,  width=7cm}}\vspace*{0.5cm}
  \centerline{\epsfig{file=fig4b.eps,  width=7cm}}

\caption{Average degree for the nearest neighbors of the vertices of
  degree $k$, $\bark(k,N)$, for the weighted growing model for $m=4$, with
  $\delta=2$ (a) and $\delta =5$ (b). Symbols correspond to the different
  system sizes $N=3 \times 10^3$ ($\circ$), $N= 10^4$ ($\Box$) $N=3 \times
  10^4$ ($\diamond$), $N=10^5$ ($\triangle$).  Top plot: Raw data.  Middle
  plot: Data rescaled by the size prefactor $N^{1-2\beta}$.  Bottom plot: Data
  rescaled by the size prefactor with logarithmic corrections.  The dashed
  lines represent a power-law decay with exponent $-2+1/ \beta$. The middle
  plots display also the values of $N^{1-2\beta} \bark^w(k,N)$ (filled
  symbols), which collapse onto a horizontal line, in agreement with the
  analytical prediction Eq.~(\protect\ref{eq:25}).}
\label{fig:knnm2d2}
\end{figure}

We have performed numerical simulations of the weighted growing
network model described in Ref.~\cite{barrat04:_weigh}, for sizes
ranging from $N=10^3$ to $N=10^5$, focusing on the behavior of the
average degree of the nearest neighbors, for both its non-weighted and
weighted versions.  In Fig.~\ref{fig:knnm2d2} we plot the average
degree of the nearest neighbors $\bark(k,N)$ for $m=2$ and $\delta=2$ (a)
which corresponds to a network with $\beta=5/6$, $\gamma=2.20$, and $\delta=5$ (b),
that yields $\beta=11/12$, $\gamma= 2.09$. As expected from the analytical
analysis performed above, the obtained scaling is analogous to the LPA
model: the numerical data follows the predicted form $\bark(k,N) \sim
N^{2\beta-1} k^{-2+1/ \beta}$.  The bottom plots highlight the presence of the
logarithmic correction of Eq.(\ref{eq:21}), by plotting the rescaled
function
\begin{equation}
  \bark^{\mathrm{resc}}(k,N) = \bark(k,N) - \frac{m}{2 \delta+1} \ln \left(\frac{2
      \delta+1}{m}\right).
\end{equation}
In this case, it is noticeable that the rescaled
$\bark^\mathrm{resc}(k,N)$ function with logarithmic corrections
yields a better data collapse than that shown by the LPA model. Even
though both models are identical at the mean field level, the existing
microscopic differences seem to yield smaller subleading corrections
for the weighted growing network model.

For this same set of parameters, we have also evaluated the weighted
average degree of the nearest neighbors, $\bark^w(k,N)$, shown in the
middle plot in Fig.~\ref{fig:knnm2d2}(a) (filled symbols). We observe
that the $\bark^w(k,N)$ is indeed, as expected, independent of $k$,
and scales with the system size with the predicted factor $N^{2 \beta
  -1}$.

\section{Conclusions}
\label{sec:conclusions}

A complete theoretical characterization of a growing network model
should imply not only the estimation of the corresponding degree
distribution, but also an analytical study of the functional form of
the correlations between the degrees of neighboring vertices.
Capitalizing on the work of Szab\'o \textit{et al.}
\cite{szabo,szaboproc}, in this paper we have provided a formalism to
compute two vertex correlations, expressed by means of the average
degree of the nearest neighbors of the vertices of degree $k$,
$\bark(k)$, valid for growing network models generated by means of the
preferential attachment mechanism and belonging to the class of the
so-called ``citation networks''. The formalism is based on a rate
equation in the continuous $k$ approximation, together with the
appropriate boundary condition, that can be easily solved in the case
in which the preferential attachment is linear in the degree.
Additionally, we have presented a more complete description of the
rate equation determining the clustering spectrum $\barc(k)$, by
discussing the effects of boundary conditions.  Applying this
framework to several growing network models, we have obtained
asymptotic expressions for the functions $\bark(k,N)$ and
$\barc(k,N)$, evidentiating both the degree dependence and the scaling
with the system size, due to finite size effects. As a general result,
we conclude that networks generated by LPA with degree exponent $\gamma<3$,
exhibit the scaling behavior
\begin{equation}
  \bark(k,N) \sim  N^{2\beta-1}k^{-2+1/\beta},
\end{equation}
previously obtained by means of scaling arguments \cite{dorogorev},
which is the signature of disassortative (negative) two vertex
correlations.  We have also been able to identify the presence of
logarithmic corrections in models with LPA, which clearly appear in
computer simulations of the model.  For this LPA model, we also
observe the presence of small assortative correlations for degree
exponents $\gamma>3$, characterized by a logarithmic growth of the
$\bark(k,N)$ function, which is otherwise independent of the network
size.  The situation is more complex in what concerns the clustering
spectrum $\barc(k,N)$. For $\gamma>3$, we observe the presence of a
crossover between two power-law decays in the degree, $\barc(k)\sim
k^{-\alpha}$, with $\alpha = -4 + 2 / \beta$ for $k \lesssim (\ln
N)^{\beta/(2\beta-1)}$, and $\alpha = -2+1/\beta$ in the asymptotic
limit, while for $\gamma>3$ we obtain an increasing $\barc(k,N)$
function, limited by an upper degree cutoff.

>From this results we can conclude that the value $\alpha \simeq 1$ observed in
the literature \cite{ravasz02,szabo} is not a generic feature of all
scale-free networks \cite{szaboproc}. However, we notice that LPA
yields networks with a vanishing clustering coefficient. In order to
assess the possible effects of this factor, we have considered the DMS
model \cite{dms}, that generates networks with a large value of $C$,
as observed in real networks. In this case, we obtain a lack of two
vertex correlations, while the clustering spectrum scales as
$\barc(k)\sim k^{-1}$. An analogous result is obtained for the similar
Holme-Kim model \cite{szabo,holme02c}. From this result we could be
tempted to conclude that a clustering spectrum scaling with exponent
$\alpha=1$ is related to the lack of correlations. However, this
explanation would be in conflict with the empirical observation of
this exponent in real networks with clear dissasortative mixing. More
work is therefore needed in order to elucidate the relations between
the scaling exponents of $\bark(k)$ and $\barc(k)$ in general complex
networks.

As a final point, we have shown the flexibility of the rate equation
approach to compute two vertex correlations by applying it to a
recently proposed weighted growing network model, in which edges are
further characterized by a distribution of weights that is dynamically
coupled to the evolving topology of the network.
%, by means of a preferential attachment mechanism. 
For this model, we are able to
extend our formalism to deal with weighted two vertex correlations,
which measure the effect of the strength of the interactions between
neighboring vertices.

The very good agreement shown between our analytical estimates and
numerical simulations suggest that the method proposed in this paper
to compute two vertex correlations is in general valid to characterize
growing citation network models.  An obvious improvement would be to
extend it to deal with models in which vertex and edge removal, and
edge rewiring, are allowed. This inclusion, however, would probably
lead to quite complex non-local rate equation, whose solution would be
much harder to tackle.

\begin{acknowledgments}
  We thank M. Alava, M. Bogu\~n\'a, and A. Vespignani for helpful comments
  and discussions. This work has been partially supported by EC-FET
  Open Project No.  IST-2001-33555 and contract 001907 (DELIS).
  R.P.-S. acknowledges financial support from the Ministerio de
  Ciencia y Tecnolog\'\i a (Spain), and from the Departament
  d'Universitats, Recerca i Societat de la Informaci\'o, Generalitat de
  Catalunya (Spain).
\end{acknowledgments}

\end{document}